\documentclass[aps,twocolumn,showpacs,preprintnumbers,prl,10pt,superscriptaddress]{revtex4-2}

\usepackage{graphicx}
\usepackage[utf8]{inputenc}
\usepackage[T1]{fontenc}
\usepackage{anyfontsize}
\usepackage{adjustbox}
\usepackage{placeins}
\usepackage{lipsum}
\usepackage{latexsym}
\usepackage{rotating}
\usepackage{multirow}
\usepackage{mathtools}
\usepackage{amsmath, amssymb, amsthm, amsfonts}
\usepackage{mathrsfs}
\usepackage{bm}
\usepackage{enumerate}
\usepackage{soul}
\usepackage{natbib}
\usepackage{verbatim}
\usepackage[normalem]{ulem}
\usepackage[dvipsnames]{xcolor}
\usepackage[unicode]{hyperref}
\hypersetup{colorlinks=true, citecolor=MidnightBlue,
            linkcolor=MidnightBlue, urlcolor=MidnightBlue, linktocpage=true}
\usepackage{orcidlink}

\newcommand{\ICASU}{Illinois Center for Advanced Studies of the Universe, Department of Physics, University of Illinois at Urbana-Champaign, Urbana, Illinois 61801, USA}
\newcommand{\CAPS}{Center for AstroPhysical Surveys, National Center for Supercomputing Applications, Urbana, Illinois 61801, USA}
\newcommand{\ASTRO}{Department of Astronomy, University of Illinois at Urbana-Champaign, Urbana, Illinois 61801, USA}
\newcommand{\SKAI}{NSF-Simons AI Institute for the Sky (SkAI), Chicago, Illinois 60611, USA}
\newcommand{\CITA}{Canadian Institute for Theoretical Astrophysics, University of Toronto, Toronto, Ontario M5S 3H8, Canada}

\newcounter{footnote_suppmat}

\begin{document}

\title{Neural Post-Einsteinian Test of General Relativity with the Third Gravitational-Wave Transient Catalog}

\author{Yiqi~Xie~\orcidlink{0000-0002-8172-577X}}
\affiliation{\ICASU}
\affiliation{\CAPS}
\affiliation{\CITA}

\author{Gautham~Narayan~\orcidlink{0000-0001-6022-0484}}
\affiliation{\ASTRO}
\affiliation{\CAPS}
\affiliation{\SKAI}

\author{Nicol\'as~Yunes~\orcidlink{0000-0001-6147-1736}}
\affiliation{\ICASU}
\affiliation{\SKAI}

\begin{abstract}
Gravitational waves (GWs) from compact binaries are excellent probes of gravity in the strong- and dynamical-field regimes.
We report a test of general relativity (GR) with the third GW Transient Catalog (GWTC-3) plus a few O4 events using the recently developed neural post-Einsteinian framework, both on individual events and at the population level through hierarchical modeling.
We find no significant violation of GR and place a constraint that, for the first time, efficiently covers non-GR theories characterized by not only post-Newtonian deviations but also those beyond under the same theory-agnostic framework.
\end{abstract}

\maketitle

\noindent{\bf{\em Introduction.}}
Having passed all experimental tests in the Solar System~\cite{Will:2014kxa} and with binary pulsars~\cite{Stairs:2003eg}, Einstein's general relativity (GR) remains our best theory for describing gravity. 
Despite that, Einstein's theory is thought to be challenged by certain theoretical issues such as the ubiquity of singularities~\cite{Penrose:1964wq,Senovilla:2014gza} and its incompatibility with quantum mechanics~\cite{Shomer:2007vq}. Moreover, GR also struggles to explain certain observed phenomena, such as the rotation curves of the galaxies~\cite{Sofue:2000jx,Bertone:2016nfn} and the late-time acceleration of the expansion rate of the universe~\cite{SupernovaSearchTeam:1998fmf,SupernovaCosmologyProject:1998vns}, without the inclusion of additional dark fields or a cosmological constant.
Considerable attention has been devoted to developing modified gravity theories, and the recent observation of gravitational waves (GWs) from compact binaries~\cite{LIGOScientific:2016aoc,LIGOScientific:2016dsl,LIGOScientific:2018mvr,LIGOScientific:2020ibl,LIGOScientific:2021usb,KAGRA:2021vkt,LIGOScientific:2024elc,LIGOScientific:2025hdt,LIGOScientific:2025slb} has opened up a new window for testing GR against these theories in the dynamical- and strong-field regimes~\cite{Berti:2018cxi}.

Given the numerous proposals for modified gravity and the computational demand of GW model building and data analysis, GW tests of GR benefit significantly from a theory-agnostic method, since the latter can lead to robust and efficient inferences about the nature of gravity. 
One of the first examples of such a theory-agnostic formalism is the parametrized post-Einsteinian (ppE) framework~\cite{Yunes:2009ke,Cornish:2011ys,Chatziioannou:2012rf,Sampson:2013lpa,Yunes:2016jcc,Tahura:2018zuq}, which constructs a metamodel for small deviations from GR through waveform amplitude and phase corrections. In the inspiral of compact binaries, the latter is prescribed through a post-Newtonian (PN) expansion\footnote{The PN formalism expands inspiral quantities in powers of $v/c$, where $v$ is the orbital velocity and $c$ is the speed of light. The expansion can be further cast into powers of GW frequency $f$ through the PN version of Kepler's third law.}~\cite{Blanchet:1995ez} with the introduction of ppE theory parameters that control the type and the magnitude of the deviation. Such a formalism has been implemented successfully both by the LIGO-Virgo-KAGRA (LVK) collaboration (in their ``parametrized inspiral test of GR''~\cite{Arun:2006yw,Mishra:2010tp,Li:2011cg,Agathos:2013upa,LIGOScientific:2016lio,Meidam:2017dgf,LIGOScientific:2018dkp,LIGOScientific:2019fpa,LIGOScientific:2020tif,LIGOScientific:2021sio,Mehta:2022pcn}), as well as by several other groups~\cite{Cornish:2011ys,Sampson:2013lpa,Nair:2019iur,Perkins:2021mhb,Perkins:2020tra,Shi:2022qno}, in parameter estimation and model selection using both synthetic and real GW data. In all such studies, a subset of the theory parameters (i.e.,~those that control the \textit{type} of GR deviation) is held constant, and parameter estimation is carried out only on the remaining ppE parameters (i.e.,~those that control the \textit{magnitude} of the GR deviation). 

The ppE approach is theory agnostic because a broad class of modifications to GR can be mapped to PN dephasings during the inspiral~\cite{Yunes:2009ke,Cornish:2011ys,Chatziioannou:2012rf,Sampson:2013lpa,Yunes:2016jcc,Tahura:2018zuq}, as long as the strength of the modification is reasonably small.
For example, scalar Gauss-Bonnet (sGB) gravity~\cite{Metsaev:1986yb,Kanti:1995vq} and dynamical Chern-Simons (dCS) gravity~\cite{Jackiw:2003pm,Alexander:2009tp} contribute to a dephasing that starts at $-1$PN order and $2$ PN order, respectively.
PpE tests (including the LVK implementation) examine GR against many theories within the above class of metamodels, taking into account only their leading PN-order contribution to the inspiral GW phase (and/or amplitude).
So far, no deviations from GR have been reported~\cite{LIGOScientific:2016lio,LIGOScientific:2018dkp,LIGOScientific:2019fpa,LIGOScientific:2020tif,LIGOScientific:2021sio,LIGOScientific:2024elc,Nair:2019iur,Perkins:2021mhb,Schumacher:2023cxh,Liu:2024atc,LIGOScientific:2025cmm}, 
and thus, constraints on certain non-GR theories can be extracted by mapping the posteriors of ppE magnitude parameters back to theory-specific parameters (e.g.,~coupling constants) (see~\cite{Nair:2019iur} for a set of examples).

Despite these constraints, the nonrejection of GR by such tests does not necessarily mean that current GW data are compatible with \textit{all} possible modifications to GR that could affect the inspiral. 
The PN description adopted by the ppE formalism presumes that the underlying non-GR effect admits a legitimate expansion in powers of the orbital velocity or the GW frequency. However, several counterexamples have recently been discovered, like the binary inspirals of compact objects that source massive scalar fields~\cite{Alsing:2011er,Berti:2012bp,Liu:2020moh,Xie:2024xex} or that have dark-photon interactions~\cite{Alexander:2018qzg,Owen:2025odr}.
In these theory-specific, non-GR examples, corrections to GR activate ``suddenly,'' and thus, they cannot be represented by a simple power law in frequency. 
Although a PN-based model may potentially detect deviations of this type~\cite{Sampson:2013jpa}, the recovery of the signal would be far from ideal, the strength of the test would be weakened, and its result would be biased to favor GR unless the signal-to-noise ratio (SNR) were unusually high or the deviation itself unusually strong.

Recently, a neural post-Einsteinian (npE) waveform model~\cite{Xie:2024ubm} was developed to mitigate the above problems for theory-agnostic inspiral tests of GR. 
Through deep learning of a variational autoencoder~\cite{1312.6114}, the npE model constructs a continuous latent space that maps dephasings from several discrete PN models. Crucially, the npE model maps non-PN dephasings to non-PN regions of the same continuous latent space. 
Additionally, this model improves the detection of PN deviations with higher PN-order corrections, and allows for a more efficient parameter estimation scheme, relative to prior implementations. 

In this Letter, we report the first data analysis application of the npE model to test GR with the third Gravitational Wave Transient Catalog (GWTC-3)~\cite{KAGRA:2021vkt} and events from the O4 discovery papers~\cite{LIGOScientific:2024elc,LIGOScientific:2025cmm,LIGOScientific:2025rsn,LIGOScientific:2025brd,LIGOScientific:2025rid}. We choose to focus on binary black hole (BBH) signals, which compose the majority of the events considered and which can be used to detect deviations in a large class of modified gravity theories, including sGB gravity, dCS gravity, Einstein-\ae ther (EA) theory~\cite{Jacobson:2000xp,Jacobson:2008aj}, khronometric gravity~\cite{Blas:2009qj,Blas:2010hb}, noncommutative gravity~\cite{Kobakhidze:2016cqh}, varying-$G$ theories~\cite{1937Natur.139..323D,Yunes:2009bv}, and theories involving massive fields such as massive sGB gravity~\cite{Yamada:2019zrb,Xie:2024xex}, which have been overlooked by previous tests. 
Hereafter, we use geometric units $G=1=c$.

\vspace{0.5em}
\noindent{\bf{\em Neural post-Einsteinian waveform.}}
The npE model for the frequency-domain inspiral signal is
\begin{align}
    \tilde{h}_{\rm npE}(f;\vec{\Xi},\vec{\zeta}) = \tilde{h}_{\rm GR}(f;\vec{\Xi})\, e^{-i\delta\Psi_{\rm npE}(f;\vec{\Xi},\vec{\zeta})}, \label{eqn:npe_wf}
\end{align}
where $\tilde{h}_{\rm GR}$ is the GR waveform\footnote{Most ppE tests, and the npE formalism, have focused on the dominant $(2,2)$ harmonic of the GW signal; other harmonics are related through a simple scaling~\cite{Chatziioannou:2012rf,Mezzasoma:2022pjb,Mehta:2022pcn}.}, which depends on source parameters $\vec{\Xi}$, such as binary masses and spins. 
Similar to the ppE metamodel, the npE waveform introduces a dephasing function, $\delta\Psi_{\rm npE}$, that additionally depends on non-GR, phenomenological parameters $\vec{\zeta}$ to capture non-GR deviations. The latter are modeled through a carefully designed and tested variational autoencoder, as described in~\cite{Xie:2024ubm}.
The difference between the ppE and the npE models is that, in the latter, the dephasing function and the parametrization are ``deeply learned'' (in a physics-informed way) to unify and extend the PN representation of the dephasing that the ppE formalism is based on. 

We follow the npE prescription of~\cite{Xie:2024ubm}, which  uses a two-dimensional npE parameter space $\vec{\zeta}=(\zeta_1,\zeta_2)$.
The npE dephasing is designed to be proportional to the polar radius and antisymmetric under $\vec{\zeta}\rightarrow-\vec{\zeta}$. 
In polar coordinates $(\zeta_b,\varphi)$, where the ``radius'' $\zeta_b$, can be negative and the angle $\varphi$ ranges within $[0,\pi)$ accordingly, the npE dephasing model is constructed as
\begin{align}
    \delta\Psi_{\rm npE}(f;\vec{\Xi},\vec{\zeta})=\zeta_b\, \kappa(\vec{\Xi},\varphi)\, \psi(Mf;\varphi), \label{eqn:npe_decomp}
\end{align}
where $M$ is the total mass of the binary.
When leading-order PN dephasings are concerned, this model automatically places each PN order along a polar line with a fixed, source-independent $\varphi$ value (through the $\psi$ function), where the PN coefficient is proportional to $\zeta_b$.
Therefore, one may interpret $\varphi$ as a generalized indicator of the non-GR theory type and $\zeta_b$ as a bilateral deviation amplitude.

To implement Eq.~\eqref{eqn:npe_decomp}, we adopt the angular function $\psi(\varphi)$ developed in~\cite{Xie:2024ubm}, which is learned by a variational autoencoder using a training set of leading PN-order dephasings, ranging from $-4$PN to $2$PN order.
This results in an ordered, almost equally spaced distribution of PN lines in a continuous angular region (and its sign-flipped image) in the $\vec{\zeta}$ space. 
We refer to the above region as ``the PN region,'' and its complement in the $\vec{\zeta}$ space as ``the non-PN region.'' 
This nomenclature is supported by the detailed parameter estimation results of~\cite{Xie:2024ubm} using simulated non-GR signals, where indeed the PN region captures dephasings that arise from a convergent PN series, and the non-PN region captures dephasings that cannot be represented as a simpler PN expansion (including nonsmooth GR deviations). 
The latter also allows the npE test to examine theories that have been overlooked by previous tests, such as sGB theory with a massive scalar field. Following~\cite{Xie:2024ubm}, we set $\varphi = 0$ at a place where $\psi$ varies most rapidly with respect to $\varphi$, i.e.,~in the middle of the non-PN region.

With $\psi(\varphi)$ determined, $\kappa(\varphi)$ is then chosen to be a positive factor in the npE model to significantly break the GR prediction outside of the unit circle $|\vec{\zeta}|=1$, so that the latter can be conveniently taken as a prior boundary for a Bayesian test of GR.
The motivation behind such a prior boundary is twofold. From a theoretical perspective, many non-GR predictions are made based on a perturbative framework, which fails when the deviation from GR becomes too large. 
From an observational perspective, the current identification of a GW signal in the detector strain relies on the assumption that the signal roughly follows the predictions of GR, and a recovery model that deviates too much from GR can risk misidentifying noise artifacts as GR deviations.

In this work, $\kappa(\varphi)$ is learned by another neural network to approximate and interpolate the following npE prior boundary at fixed PN angles [defined by the already-learned $\psi(\varphi)$ function]:
\begin{align}
    \mathcal{N}^2[\delta\Psi_{\rm npE}(\vec{\Xi},\vec{\zeta})]\Big|_{|\vec{\zeta}|=1}=\mathcal{N}^2[\Psi_{\rm GR}^{\rm 0PN}(\vec{\Xi})],
\end{align}
where $\Psi_{\rm GR}^{\rm 0PN}$ is the leading PN-order GW phase in GR, and
\begin{align}
    \mathcal{N}^2[\Psi]=\int\frac{|\tilde{h}(f)|^2\Psi(f)^2}{4\pi^2{\rm SNR}^2 S_n(f)}\, df
\end{align}
is a measure inspired by effective cycles~\cite{Sampson:2014qqa}, which estimate the number of GW cycles incurred by the phasing function $\Psi$ as weighted by the noise power spectral density $S_n$.
Here, we choose $S_n$ as an average estimate during LVK's third observing run~\cite{KAGRA:2013rdx}\footnote{\url{https://dcc.ligo.org/LIGO-T2000012/public}}, from which most data for our test are collected. 
Note that the above prescription for $\kappa$ is similar to but not exactly the same as that in~\cite{Xie:2024ubm}, and we elaborate more on the difference in the Supplemental Material~\footnote{See Supplemental Material, which includes Refs.~\cite{Zhong:2024pwb,Mathematica,2020SciPy-NMeth,scott2015multivariate}, for details about the npE parametrization of the waveform deviation, hierarchical modeling for the combined npE test, computational settings for the data analysis, posterior results from individual GW events with the diagnosis of noise artifacts, and posterior results from the combined npE test using the hierarchical model.}\setcounter{footnote_suppmat}{\value{footnote}}.

\vspace{0.5em}
\noindent{\bf{\em Gravitational wave parameter estimation.}}
We use LVK open data~\cite{LIGOScientific:2019lzm,KAGRA:2023pio} and focus on events selected for the LVK parametrized inspiral tests of GR~\cite{LIGOScientific:2020tif,LIGOScientific:2021sio}, each of which is (i) detected by at least two detectors, (ii) has a false-alarm rate (FAR) less than $10^{-3}\,{\rm yr}^{-1}$, and (iii) accumulates an SNR greater than $6$ during the inspiral. 
We further filter the list by requiring that the sources have been confirmed as BBHs. 
We additionally consider GW241011\_233834 and GW250114\_082203, featured as multidetector, low-FAR, and inspiral-loud BBH events from the O4 discovery papers~\cite{LIGOScientific:2024elc,LIGOScientific:2025cmm,LIGOScientific:2025rsn,LIGOScientific:2025brd,LIGOScientific:2025rid}.
This leaves us with 27 events (see Supplemental Material~\footnotemark[\value{footnote_suppmat}] for a full list).

For each event, we perform Bayesian parameter estimation with the waveform model of Eq.~\eqref{eqn:npe_wf} and a Gaussian noise model. 
The $\tilde{h}_{\rm GR}$ function in Eq.~\eqref{eqn:npe_wf} is taken to be \texttt{IMRPhenomPv2}~\cite{Hannam:2013oca,Husa:2015iqa,Khan:2015jqa} by default, which well models the $(2,2)$ GW mode as the dominant signal from a symmetric BBH.
This leaves GW190412~\cite{LIGOScientific:2020stg} as a special case in our selection, given its observational evidence for significant higher-multipole modes~\cite{LIGOScientific:2020stg}. For this event, we use \texttt{IMRPhenomXPHM}~\cite{Pratten:2020ceb,Pratten:2020fqn,Garcia-Quiros:2020qpx} with an additional $(3,3)$ mode that reasonably recovers the remaining SNR beyond the $(2,2)$ mode.

Exploiting the \texttt{Bilby} inference library~\cite{Ashton:2018jfp} with the \texttt{dynesty} nested sampler~\cite{Speagle:2019ivv}, we estimate the posterior distribution for $\vec{\Xi}$ and $\vec{\zeta}$ and marginalize over the former.
The prior choice for $\vec{\Xi}$ is adapted from the LVK standard analysis assuming GR~\cite{LIGOScientific:2018mvr,LIGOScientific:2020ibl,LIGOScientific:2021usb,KAGRA:2021vkt,Planck:2015fie}. 
In the npE sector, we consider both $(\zeta_1,\zeta_2)$ and $(\zeta_b,\varphi)$ for parametrizing the model and, in each case, we choose a uniform prior over the two parameters within the unit circle.

For each event, we repeat the above parameter estimation for the strain data from each individual detector. We find three events for which the individual-detector posteriors appear to be incompatible with each other, implying that the npE tests on these events are likely impacted by detector-specific noise artifacts. 
One of the three events affected is GW200129\_065458, which is known to have a glitch-removal artifact in the LVK open data that can impact the inference of spin precession~\cite{Hannam:2021pit,Payne:2022spz,Macas:2023wiw,Estelles:2025zah}, eccentricity~\cite{Gupte:2024jfe,Planas:2025jny,Chiaramello:2025bhi}, and GR deviations~\cite{Maggio:2023vch,Gupta:2024gun} (see~\cite{Macas:2023zdu,Macas:2023wiw} for more on the glitch mitigation of this event). The method of cross-checking posteriors obtained with different networks for the same event has been useful in identifying anomalies associated with noise features (see, e.g.,~\cite{Ghosh:2023mdo,Ghosh:2025axc,Payne:2022spz}).  
We exclude the above three events from the results presented hereafter.

To make full use of the remaining events, we introduce a hierarchical model to combine the marginalized $\vec{\zeta}$ posteriors from individual-event npE tests. 
Similar to the hierarchical model~\cite{Zimmerman:2019wzo,Isi:2019asy,Isi:2022cii} employed in the LVK analysis~\cite{LIGOScientific:2020tif,LIGOScientific:2021sio}, we assume that the bilateral deviation $\zeta_b$ follows a Gaussian distribution with a mean $\mu$ and a standard deviation $\sigma$. On the other hand, the polar angle $\varphi$ represents the type of theory deviation, and thus, it can be kept constant, as it must be shared across all events (i.e.,~we assume that any modification to GR impacts all BBH events detected and not only a subset).
We apply our model to GWTC-3, excluding the two O4 events, so that we can compare with LVK's test results from the same catalog.
We neglect the selection effect in our model as its impact on theory-agnostic inspiral tests of GR is small given current data~\cite{Payne:2023kwj,Magee:2023muf}.

For the hierarchical inference, we choose a prior uniform over $\mu\in[-1,1]$, $\sigma\in[0,1]$, and $\varphi\in[0,\pi]$. 
We customize \texttt{Bilby} to sample over these parameters, where we reuse the individual-event npE posterior samples to compute the hierarchical likelihood. 
Once the hyperparameters $\{\mu,\sigma,\varphi\}$ are estimated, we extract the posterior quantile for GR ($\mu=0=\sigma$) and reconstruct the $\vec{\zeta}$ population for comparison with LVK test results. 
See the Supplemental Material~\footnotemark[\value{footnote_suppmat}] for the detailed settings of our individual-event parameter estimation and hierarchical inference. 

\begin{figure}[htbp]
    \centering
    \includegraphics[width=0.95\linewidth]{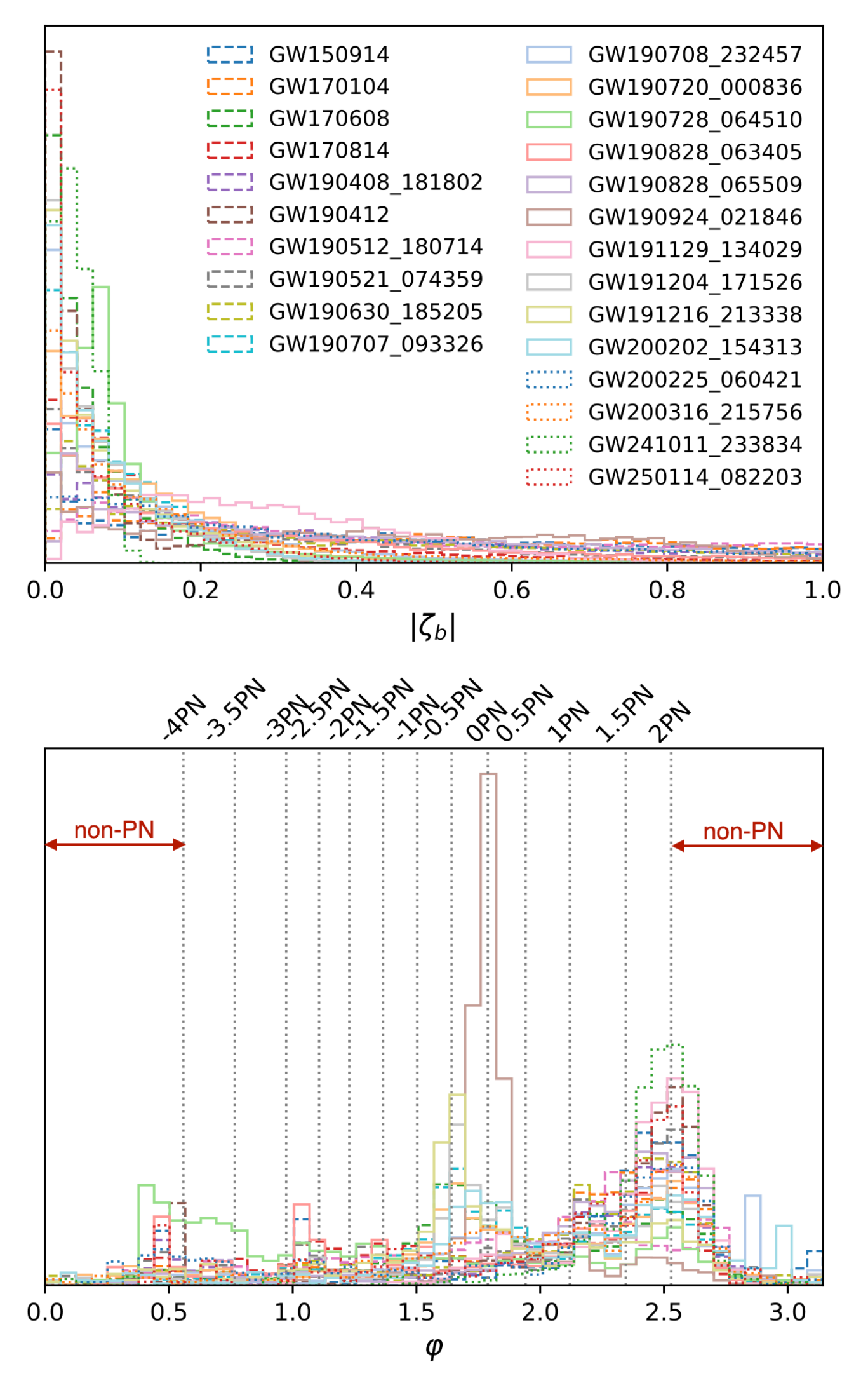}
    \caption{
    Individual-event marginalized posteriors of the npE deviation $|\zeta_b|$ (upper panel) and the theory angle $\varphi$ (lower panel). Observe that all $|\zeta_b|$ posteriors are attached with GR at $\zeta_b=0$, suggesting no significant deviation from GR in general. For $\varphi$, gray dotted lines are added to label the directions of PN dephasings. Observe that the $\varphi$ posteriors mostly peak around the $0$PN direction and the $2$PN direction, which reflects the expected correlation between the npE parameters and the GR parameters through their mutual contribution to the GW phase at these PN orders. Apart from that, no significant preference is found toward those non-PN theory angles (to the left of the $-4$PN line or to the right of the $2$PN line).}
    \label{fig:individual_event_summary}
\end{figure}

\vspace{0.5em}
\noindent{\bf{\em Constraints on non-GR deviations.}}
Figure~\ref{fig:individual_event_summary} presents the individual-event marginalized posterior of $|\zeta_b|$ and $\varphi$ estimated using uniform priors over $\zeta_b$ and $\varphi$. The $|\zeta_b|$ results suggest that all events are compatible with GR, except that the support from GW191129\_134029 is marginal, which may be attributed to instrument noise (see a similar case in~\cite{LIGOScientific:2016lio}). 
Furthermore, from the $\varphi$ plot, we see no preference toward the non-PN region, which complements previous conclusions drawn by the ppE LVK test. 
Within the PN region, most events select $\varphi$ values toward the $2$PN end. 
This is because the positive PN-order deviations are more correlated with existing GR parameters, such as the spins and mass ratio, whose effects also take place at positive PN orders. In addition, observe that a few events strongly prefer the $0$PN angle, which is due to the correlation with the chirp mass (as first found in~\cite{Cornish:2011ys}), which accounts for the dominating effect of quadrupole radiation.

\begin{figure}[htbp]
    \centering
    \includegraphics[width=0.95\linewidth]{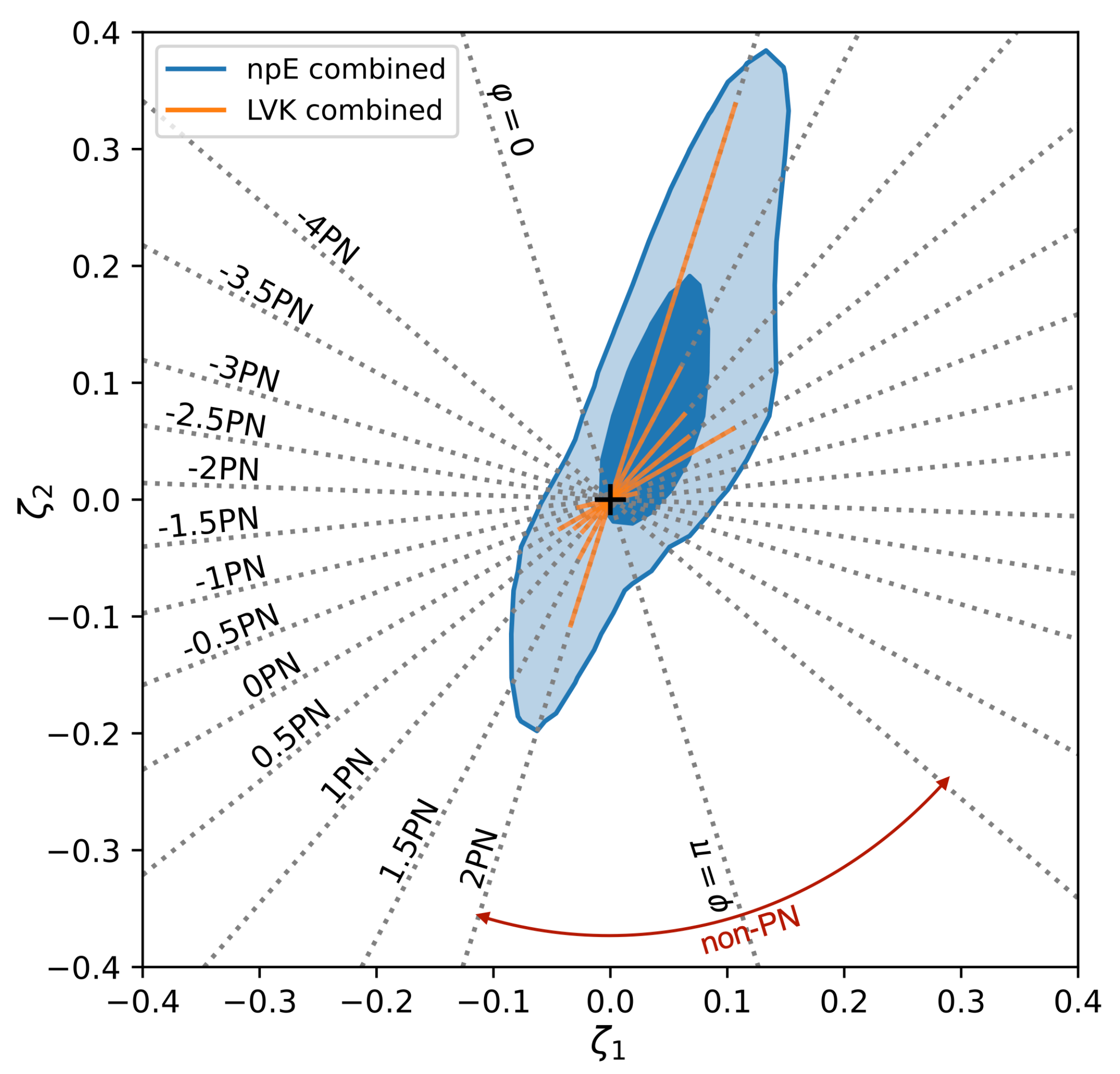}
    \caption{
    Combined GWTC-3 npE constraint using a hierarchical model. The blue contours enclose the 50\% and 90\% credible regions of a $\vec{\zeta}$ distribution reconstructed from the posterior of the hierarchical inference. 
    The gray dotted lines mark special directions as annotated. Apart from the same PN lines in Fig.~\ref{fig:individual_event_summary}, we also show the angles where $\varphi$ is defined to be $0$ and $\pi$.
    The black ``+'' at the center marks GR.
    For comparison, we take the LVK posteriors published in~\cite{LIGOScientific:2019fpa,LIGOScientific:2020tif,LIGOScientific:2021sio} and overlay their 90\% credible intervals in the npE parameter space as orange lines, whenever the mapping is applicable. 
    Observe that our combined npE constraint is compatible with GR and roughly reproduces the LVK results. Moreover, the npE constraint suggests no significant deviation from GR in the non-PN region, as well as the area between integer and half-integer PN orders, where higher PN-order corrections to GR deviations reside~\cite{Xie:2024ubm}.
    }
    \label{fig:population}
\end{figure}

Figure~\ref{fig:population} presents the reconstructed $\vec{\zeta}$ population from the hierarchical npE test. Observe that the population distribution dies off well within the npE prior boundary $|\vec{\zeta}|=1$, which justifies our reuse of the individual-event posteriors for the hierarchical estimation, despite the fact that a few individual-event posteriors (e.g.,~the GW190512\_180714 posterior in Fig.~\ref{fig:individual_event_summary}) do not die off sufficiently fast at the same boundary.

The combined npE constraint on non-GR deviations is more straightforwardly given by the 90\% credible contour extracted from the hierarchically reconstructed $\vec{\zeta}$ population. 
Similar to the observation from the individual-event posteriors, we find no evidence for non-PN deviations. The constraint loosens near the $2$PN line, approaching $|\vec{\zeta}|\lesssim0.4$ at maximum. 
Moreover, the underlying estimation of the hyperparameters $(\mu,\sigma,\varphi)$ suggests a GR quantile of $Q_{\rm GR}=0.34$, i.e.,~GR falls at the $34\%$ quantile of the hyperparameters posterior.

To compare our constraint with the LVK's, we take the GWTC-3 combined posteriors of LVK ppE deviations along $-1$PN, $0$PN, and positive PN orders (estimated by freeing one PN coefficient at a time)~\cite{LIGOScientific:2021sio,ligo_scientific_collaboration_and_virgo_2023_8177023,ligo_scientific_collaboration_and_virgo_2022_6513631,ligo_scientific_collaboration_virgo_coll_2022_7007370} and map their 90\% credible intervals to the npE parameter space. 
Observe that our hierarchical npE 90\% credible contour roughly reproduces the LVK results whenever the latter are within the npE PN range, although the npE constraint tends to be more conservative due to internal correlations~\cite{Xie:2024ubm} and the fact that the LVK analysis considered more events including a couple of neutron star black holes.
Unlike standard ppE tests, including the LVK's, the npE constraint covers the non-PN region as well as the area ``between'' integer and half-integer PN orders, where higher PN-order corrections to GR deviations reside~\cite{Xie:2024ubm}.
Thus, the npE constraint extends the LVK results and leads to a more robust test of GR.

\vspace{0.5em}
\noindent{\bf{\em Discussion and future prospects.}}
We have conducted the first npE test of GR using BBHs' inspirals from GWTC-3 and O4 discovery papers, where we investigate deviations from GR under a theory-agnostic parametrization for both individual events and the combined population across GWTC-3 using a customized hierarchical model. 
We find that the data do not support any significant deviation from GR, and thus, we place the first constraint on non-GR deviations covered within the npE parameter space. 
These deviations include PN dephasings from the GR signal with leading PN orders ranging from $-4$PN to $2$PN, covering a broad class of theories that include dCS gravity, sGB gravity, EA gravity, khronometric gravity, noncommutative gravity, and varying-$G$ gravity.

In addition, the npE parameter space also contains a non-PN region for capturing deviations that cannot be described by smooth PN dephasings.
These deviations can be motivated by theories in which the binary system is coupled to auxiliary massive fields.
Reference~\cite{Xie:2024xex} searched for dipole emission from massive scalar fields, such as that which arises in massive sGB theory, using LVK BH binaries, and the search returned a null detection.
Our results confirm the above conclusion under a more agnostic framework, with broader implications potentially covering vector fields and Yukawa forces in the conservative sector of the orbital dynamics~\cite{Alexander:2018qzg,Zhang:2021mks,Owen:2025odr,Xie:2024ubm}. 

The current npE waveform model is built on neural networks trained with BBH signals, and in this work, we only (conservatively) apply the npE test to BBH events. This means we cannot make any inferences on non-GR theories that do not modify BBH signals, such as scalar-tensor theories, like Brans-Dicke theory~\cite{Sotiriou:2008rp,Kobayashi:2019hrl} and theories with dark-photon interactions in the hidden sector~\cite{Alexander:2018qzg,Owen:2025odr}. On the other hand, these theories may leave imprints when the binary involves at least one NS, and there is an ongoing effort to upgrade the npE model so that it can be effectively applied to NS binaries~\cite{Loane:inprep}. Since BBHs compose the majority of the GWTC-3 sources, our results use the most information available while still covering possible deviations that arise from a wide set of theories.

\vspace{0.5em}
\begin{acknowledgments}
\noindent{\bf{\em Acknowledgments.}}
Y.X. and N.Y. acknowledge support from the Simons Foundation through Award No.~896696, the Simons Foundation International through Award No.~SFI-MPS-BH-00012593-01, the NSF through Grants No.~PHY-2207650 and No.~PHY-25-12423, and NASA through Grant No.~80NSSC22K0806. 
Y.X. also acknowledges support from the Illinois Center for Advanced Studies of the Universe (ICASU)/Center for AstroPhysical Surveys (CAPS) Graduate Fellowship and support of the Natural Sciences and Engineering Research Council of Canada (NSERC) [funding reference number 513671].
G.~N. acknowledges NSF support from AST-2206195, and a CAREER grant, supported in part by funding from Charles Simonyi, NSF AST 2421845 and support from the Simons Foundation as part of the NSF-Simons SkAI Institute. This work made use of the Illinois Campus Cluster, a computing resource that is operated by the Illinois Campus Cluster Program (ICCP) in conjunction with the National Center for Supercomputing Applications (NCSA), and is supported by funds from the University of Illinois Urbana-Champaign (UIUC).
\end{acknowledgments}

\noindent{\bf{\em Data availability}.}
The data that support the findings of this article are openly available~\cite{LIGOScientific:2019lzm,KAGRA:2023pio,ligo_scientific_collaboration_and_virgo_2023_8177023,ligo_scientific_collaboration_and_virgo_2022_6513631,ligo_scientific_collaboration_virgo_coll_2022_7007370}.

\bibliography{references.bib}

\clearpage
\appendix
\section*{Supplemental Material}

\subsection{Customization and change of notation of the npE waveform model}
The npE dephasing in the main text $\delta\Psi_{\rm npE}(f;\vec{\Xi},\vec{\zeta})$ is customized based on the one developed in~\cite{Xie:2024ubm}, but we here use different notation in some parts of the model for readability. We describe below the differences in detail, beginning with a brief review of the original npE model in~\cite{Xie:2024ubm}.

In~\cite{Xie:2024ubm}, the original npE dephasing was designed as
\begin{align}
    \delta\Phi_{\rm npE}(f;\vec{\Xi},\vec{z})=\lVert\vec{z}\rVert\,T(\vec{\Xi},\hat{z})\,S(Mf;\hat{z}), \label{eqn:npe_original}
\end{align}
where $\vec{z}$ is a set of non-GR parameters, $\lVert\cdot\rVert$ refers to the $L_2$ norm, $\hat{(\cdot)}$ refers to the $L_2$ normalized version (or the direction) of a vector, $S$ is called the ``shape function'' because it controls how the dephasing varies over frequencies, and $T$ is called the ``scale function'' because it controls the dephasing magnitude. 
The actual functional forms of $S$ and $T$ are determined each by different neural networks on a discrete set of $Mf$ values, where $S$ and $T$ are enforced to be antisymmetric and symmetric, respectively, under $\hat{z}\rightarrow-\hat{z}$. Additionally, $T$ is enforced to be positive definite.

The shape function $S$ is first learned by a variational autoencoder to distribute a training set of PN dephasings across $\hat{z}$ directions. This is a semi-supervised learning process, i.e.~the neural network is only given a variety of shapes as the learning material, but the parametrization of $S$ with $\hat{z}$ is generated by the network itself during the process without any guidance from the training set. 
After training, the network identifies a set of directions $\hat{z}_{(i/2){\rm PN}}$, where the $(i/2)$PN power laws $(\pi Mf)^{(-5+i)/3}$ are reproduced (up to a scale factor) by the $S$ output.

Once the shape function $S$ is determined, the scale function $T$ is learned by a secondary network (with weights in the shape network frozen) to interpolate the prior boundary, such that along the $(i/2)$PN direction in the $\vec{z}$ space, we have 
\begin{align}
    \delta\Phi_{\rm npE}(f;\vec{\Xi},\hat{z}_{(i/2){\rm PN}})\approx p_i(\vec{\Xi})(\pi Mf)^{(-5+i)/3}, \label{eqn:npe_pn_match}
\end{align}
where the approximation is approached by minimizing a loss function, and
\begin{align}
    p_i(\vec{\Xi})=\left\{\begin{aligned}
        &|p_i^{\rm GR}(\vec{\Xi})|,&\,&i=0~{\rm or}~i\geq2, \\
        &|p_0^{\rm GR}(\vec{\Xi})p_2^{\rm GR}(\vec{\Xi})|^{1/2},&\,&i=1, \\
        &|p_0^{\rm GR}(\vec{\Xi})|\,(\pi Mf_{\rm low})^{-i/3},&\,&i<0,
    \end{aligned}\right. 
\end{align}
where $p_i^{\rm GR}$ is the $(i/2)$PN coefficient of the GR phase and $f_{\rm low}$ is the lower frequency bound of the detector sensitivity band. 
The $p_i$ function extends the PN coefficient in GR when the latter becomes identically zero, and Eq.~\eqref{eqn:npe_pn_match} essentially leads to a prior boundary at which the npE dephasing saturates a 100\% fractional deviation from GR as measured by the effective PN coefficient.

In~\cite{Xie:2024ubm}, one realization of the above design has been obtained based on a two-dimensional representation of $\vec{z}$ and using a training dataset that included integer-PN-order dephasings between $-4$PN and $2$PN for a population of BBHs (hence, a population of $p_i(\vec{\Xi})$). 
This resulted in two trained neural networks: one for the shape function and another for the scale function.
In this paper, we follow the same decomposition of Eq.~\eqref{eqn:npe_original} (as one can see from the correspondence between $\vec{z}$--$\vec{\zeta}$, $T$--$\kappa$, and $S$--$\psi$) and partially inherit the previously developed networks. However, we do implement several changes that we detail below. 

Let us begin by discussing the shape network. We do adopt the same network as in~\cite{Xie:2024ubm}, so that the polar angle $\varphi$ in the $\vec{\zeta}$ space of the main text is the same as the polar angle for $\vec{z}$ in~\cite{Xie:2024ubm}. 
However, for readability, we do not introduce the notation $\lVert\cdot\rVert$ and $\hat{(\cdot)}$ of the main text. Instead, we only formally describe the npE dephasing model with shape function $\psi(\varphi)$ in place of $S(\hat{z})$, and we note that they encode the same shape information as
\begin{align}
    {\rm sign}(\zeta_b)\psi(\varphi)=S(\hat{z}).
\end{align}

Let us now discuss the scale network. In this case, we only adopt the architecture of~\cite{Xie:2024ubm} (i.e.~the depth and width of the network, the type of neural activation functions, the way different layers get connected, etc.), but we redo the training of the network, and we define the prior boundary differently, based on an effective-cycles criterion.
In particular, we retrain the scale network to approximate
\begin{align}
    \delta\Psi_{\rm npE}(f;\vec{\Xi},\hat{\zeta}_{(i/2){\rm PN}}) \approx q_i(\vec{\Xi})(\pi Mf)^{(-5+i)/3},
\end{align}
where
\begin{align}
    q_i(\vec{\Xi})=\sqrt{\frac{\mathcal{N}^2[\Psi_{\rm GR}^{\rm 0PN}(\vec{\Xi})]}{\mathcal{N}^2[(\pi Mf)^{(-5+i)/3}]}}.
\end{align}
More specifically, we retrain the scale network with a new training dataset created from $q_i(\vec{\Xi})$ instead of $p_i(\vec{\Xi})$. The loss function and the training procedure, however, follow the previous prescription of~\cite{Xie:2024ubm}, and we have verified that the same recipe still leads to good convergence at the end of the training process. 
We denote the new scale function via $\kappa$, and we note that $\vec{z}$ and $\vec{\zeta}$ differ only by how their polar radii are mapped to the magnitude of the npE dephasing.

To summarize, the npE dephasing $\delta\Psi_{\rm npE}$ in this work is related to the original $\delta\Phi_{\rm npE}$ as
\begin{align}
    \delta\Psi_{\rm npE}(f;\vec{\Xi},\vec{\zeta}=\vec{z})\propto \delta\Phi_{\rm npE}(f;\vec{\Xi},\vec{z}),
\end{align}
where the coefficient of proportionality depends only on $\vec{\Xi}$ and $\varphi$ (or $\hat{z}$).
The two dephasing functions share the same shape across frequencies and differ only by their overall scales.
In this regard, the distribution of theory types across the npE polar angle (including the positioning of the PN lines) in this work is exactly the same as that in~\cite{Xie:2024ubm}, but the magnitude of the deviations decoded from the npE polar radii differs, such that the new npE prior boundary at the unit circle complies with the effective-cycles criterion. 

\subsection{Hierarchical model for npE deviations}
In the main text, we combine npE test results across individual events to make full use of the catalog. 
Here, the assumption is that there exists one unique theory (with a unique set of coupling constants) behind all GW signals observed, and the measured $\vec{\zeta}$ values from different events must collectively follow a certain population, as predicted by that theory and by astrophysics. 
Therefore, the goal is to extract the $\vec{\zeta}$ population from individual observations and compare it with the GR prediction. 
Under a Bayesian framework, this can be tackled through hierarchical inference, using a parametrized model for the $\vec{\zeta}$ population.

Because we aim for a theory-agnostic test, the hierarchical model must be generic. For ppE tests, a well-justified hierarchical model has been proposed in~\cite{Isi:2019asy} and widely applied in various LVK analyses~\cite{LIGOScientific:2020tif,LIGOScientific:2021sio}. In such a model, the ppE deviation parameter at each PN order is assumed to follow a Gaussian population with a certain mean and standard deviation. 
In this work, we extend the above design and assume the following model for the npE population distribution
\begin{align}
    p_{\rm pop}(\zeta_b,\varphi|\mu,\sigma,\bar{\varphi}) = \frac{1}{\sqrt{2\pi}\sigma} e^{-\frac{(\zeta_b-\mu)^2}{2\sigma^2}} \delta(\varphi-\bar{\varphi}). \label{eqn:npe_pop_model}
\end{align}
We here assume that the bilateral deviation $\zeta_b$ follows a Gaussian distribution with mean $\mu$ and standard deviation $\sigma$. The theory angle $\varphi$, however, is shared across all events\footnote{In this Supplemental Material, we use the barred symbol $\bar{\varphi}$ to denote the theory angle of the population model. This is slightly different from the presentation in the main text, where we reused $\varphi$ for simplicity. The choice here is made for a clearer distinction between the npE parameter and the population hyperparameter.}. 
We note that our prescription is different from that in~\cite{Zhong:2024pwb}, where a similar test of GR was considered, and a multivariate Gaussian distribution was proposed for modeling the population of more than one deviation parameter. In our prescription, $\varphi$ is chosen to be fixed because it has the interpretation of a (universal) theory type. 

According to our population model, GR corresponds to $\mu=0=\sigma$, i.e.~the npE deviation is always zero, and a parametrized test of GR can be constructed by examining the above as a null hypothesis against data from the GW catalog.
In order to do this, we estimate the population parameters using the following hierarchical likelihood
\begin{align}
    \mathcal{L}_{\rm h}(\{s\}|\mu,\sigma,\bar{\varphi})&=\;\int \prod_{i=1}^{N} \mathcal{L}_{\rm npE}^{(i)}(s^{(i)}|\zeta_b^{(i)},\varphi^{(i)})\, 
    \nonumber \\
    &\times p_{\rm pop}(\zeta_b^{(i)},\varphi^{(i)}|\mu,\sigma,\bar{\varphi})\, d\zeta_b^{(i)}\,d\varphi^{(i)}, \label{eqn:likelihood_hierarchical}
\end{align}
where $\{s\}\equiv\{s^{(1)},s^{(2)},\cdots,s^{(N)}\}$ is the set of strain data from $N$ events in the catalog, and $\mathcal{L}_{\rm npE}^{(i)}$ is the individual-event GW likelihood, obtained with the npE waveform model (see~\cite{Xie:2024ubm} for details). 

With priors specified, Bayes' theorem can be applied to the hierarchical likelihood $\mathcal{L}_{\rm h}$ to obtain the posterior distribution for the hyperparameters, $p_{\rm h}(\mu,\sigma,\bar{\varphi}|\{s\})$, based on which the GR quantile can be introduced as~\cite{Isi:2022cii},
\begin{align}
    Q_{\rm GR}=\int_{p_{\rm h}(\mu,\sigma|\{s\})>p_{\rm h}(0,0|\{s\})}p_{\rm h}(\mu,\sigma|\{s\})\, d\mu\,d\sigma.
    \label{eqn:gr_quantile}
\end{align}
Here, $p_{\rm h}(\mu,\sigma|\{s\})$ is the posterior after marginalizing over $\bar{\varphi}$. 
The GR quantile measures how much GR ($\mu=0=\sigma$) is disfavored by the hierarchical inference, with $Q_{\rm GR}=0$ placing GR at the posterior peak and $Q_{\rm GR}=1$ excluding GR from any support of the posterior. 
The hierarchical posterior can also lead to a reconstructed population distribution~\cite{Isi:2019asy},
\begin{align}
    p_{\rm recon}(\zeta_b,\varphi|\{s\})=&\;\int  p_{\rm pop}(\zeta_b,\varphi|\mu,\sigma,\bar{\varphi}) \notag\\
    &\quad\times p_{\rm h}(\mu,\sigma,\bar{\varphi}|\{s\})\, d\mu\,d\sigma\,d\bar{\varphi},
    \label{eqn:pop_recon}
\end{align}
from which the ``combined constraint'' in the main text (as shown in Fig.~2) can be extracted.

As a final remark, the numerical evaluation of Eq.~\eqref{eqn:likelihood_hierarchical} can be greatly simplified when each individual-event likelihood $\mathcal{L}_{\rm npE}$ is represented as a sum of Gaussian density functions of $\vec{\zeta}$. Consider, for example, 
\begin{align}
    \mathcal{L}_{\rm npE}(s|\vec{\zeta}) = \sum_{j=1}^K \frac{w_j}{\sqrt{2\pi|C_j|}} e^{-\frac{1}{2}(\vec{\zeta}-\vec{\mu}_j)^TC_j^{-1}(\vec{\zeta}-\vec{\mu}_j)},
\end{align}
where the likelihood has been decomposed into a sum of $K$ Gaussian density functions, and for the $j$th component, $w_j$ is the $j$th weight, $\vec{\mu}_j$ is the $j$th Gaussian mean, and $C_j$ is the $j$th covariance matrix, with $|C_j|$ its determinant.
Using $\vec{\zeta}=(\zeta_b\cos\varphi,\zeta_b\sin\varphi)$, and the fact that $p_{\rm pop}$ contains another Gaussian density function, the integral in Eq.~\eqref{eqn:likelihood_hierarchical} can be analytically solved to obtain
\begin{widetext}
\begin{align}
    \mathcal{L}_{\rm h}(\{s\}|\mu,\sigma,\bar{\varphi}) = 
    \prod_{i=1}^N \sum_{j=1}^{K^{(i)}} \Bigg\{ &\; \frac{w_j^{(i)}}{2\pi\sqrt{\big|C_j^{(i)}\big|\big(1+\sigma^2\,\vec{n}_{\bar{\varphi}}^T {\big(C_j^{(i)}\big)}^{-1} \vec{n}_{\bar{\varphi}}\big)}} \notag\\
    &\;\times \exp\Bigg[ -\frac{1}{2\sigma^2} \Bigg( \mu^2 + \sigma^2 {\vec{\mu}_j^{(i)T}} {\big(C_j^{(i)}\big)}^{-1} \vec{\mu}_j^{(i)} - \frac{\mu+\sigma^2 {\vec{\mu}_j^{(i)T}} {\big(C_j^{(i)}\big)}^{-1} \vec{n}_{\bar{\varphi}}}{1+\sigma^2\,\vec{n}_{\bar{\varphi}}^T {\big(C_j^{(i)}\big)}^{-1} \vec{n}_{\bar{\varphi}}} \Bigg) \Bigg] \Bigg\}, \label{eqn:likelihood_hierarchical_gaussian}
\end{align}
\end{widetext}
where $\vec{n}_{\bar{\varphi}}=(\cos\bar{\varphi},\sin\bar{\varphi})$. We have verified the above solution using \texttt{Mathematica}~\cite{Mathematica}.

\subsection{Computational settings}
The events analyzed in this work are explicitly listed in Table~\ref{tab:events}.
As pointed out in the main text, in addition to the filter applied by the LVK parametrized inspiral tests~\cite{LIGOScientific:2020tif,LIGOScientific:2021sio}, we further require that the source be confidently identified as a BBH, which has eliminated possible NSBH events, such as GW190814.
We load strain data from the Gravitational Wave Open Science Center~\cite{LIGOScientific:2019lzm,KAGRA:2023pio}, and follow the same choice of signal duration, frequency range, noise spectral density estimates and glitch mitigation as that described in~\cite{LIGOScientific:2018mvr,LIGOScientific:2020ibl,LIGOScientific:2021usb}. 

\begingroup
\setlength{\tabcolsep}{5pt}
\begin{table}[htbp]
    \centering
    \begin{tabular}{l|l|l}
        \hline
        Event identifier & Detectors & Note \\
        \hline
        GW150914 & HL & -- \\
        GW151226 & HL & Noise artifact \\
        GW170104 & HL & -- \\
        GW170608 & HL & -- \\
        GW170814 & HLV & -- \\
        GW190408\_181802 & HLV & -- \\
        GW190412 & HLV & Higher harmonics~\cite{LIGOScientific:2020stg} \\
        GW190512\_180714 & HLV & -- \\
        GW190521\_074359 & HL & -- \\
        GW190630\_185205 & HLV & -- \\
        GW190707\_093326 & HL & -- \\
        GW190708\_232457 & LV & -- \\
        GW190720\_000836 & HLV & -- \\
        GW190728\_064510 & HLV & -- \\
        GW190828\_063405 & HLV & -- \\
        GW190828\_065509 & HLV & -- \\
        GW190924\_021846 & HLV & -- \\
        GW191129\_134029 & HL & -- \\
        GW191204\_171526 & HL & -- \\
        GW191216\_213338 & HV & -- \\
        GW200129\_065458 & HLV & Noise artifact \\
        GW200202\_154313 & HLV & -- \\
        GW200225\_060421 & HL & -- \\
        GW200311\_115853 & HLV & Noise artifact \\
        GW200316\_215756 & HLV & -- \\
        GW241011\_233834 & HV & Higher harmonics~\cite{LIGOScientific:2025brd} \\
        GW250114\_082203 & HL & -- \\
        \hline
    \end{tabular}
    \caption{Events selected for our analysis and the list of detectors operated during each event. For the latter, the abbreviations ``H,'' ``L,'' and ``V'' correspond to the Hanford detector, the Livingston detector, and the Virgo detector, respectively.
    As pointed out later, three events fail the consistency check when comparing the npE posteriors from different detectors, suggesting certain noise artifacts that may be significantly affecting the inference process. These events are removed from the results presented in the main text. In addition, the strain data for GW190412 and GW250114\_082203 is known to have significant contributions from higher harmonics, and so we take special care when modeling these signals. 
    }
    \label{tab:events}
\end{table}
\endgroup

By default, we choose \texttt{IMRPhenomPv2} as the base GR waveform $\tilde{h}_{\rm GR}$. However, in the special cases of GW190412 and GW250114\_082203, we choose \texttt{IMRPhenomXPHM} with an additional $(3,3)$ mode added on top of the dominant $(2,2)$ mode, as mentioned in the caption of Table~\ref{tab:events}. Both GR waveforms are parametrized by
\begin{align}
    \vec{\lambda}_{\rm GR} = \{m_1, m_2, \vec{\chi}_1, \vec{\chi}_2, t_c, \phi_{\rm ref}, \psi, \iota, \alpha, \delta, D_L\},
\end{align}
where $m_{1,2}$ are the component masses, $\vec{\chi}_{1,2}$ are the component dimensionless spin vectors, $t_c$ is the coalescence time, $\phi_{\rm ref}$ is a reference phase, $\psi$ is the polarization angle, $\iota$ is the inclination angle, $\alpha$ is the angle of right ascension, $\delta$ is the declination angle, and $D_L$ is the luminosity distance. 

Similar to the LVK analysis of~\cite{LIGOScientific:2018mvr,LIGOScientific:2020ibl,LIGOScientific:2021usb,KAGRA:2021vkt}, we choose a uniform prior over the redshifted component masses, spin magnitudes, coalescence time and reference phase, and an isotropic prior over the spin orientation, binary orientation and sky location. 
In particular, the prior over the masses is restricted to $m_2/m_1\in[0.125,1]$ for \texttt{IMRPhenomPv2} and $[0.05,1]$ for \texttt{IMRPhenomXPHM}. 
The prior over the spin magnitudes ranges inside $[0,0.99]$.
The prior over the coalescence time is restricted to $\pm0.1\,{\rm s}$ around the trigger time of the event.
For the luminosity distance, we choose a prior that is uniform in the source frame volume. A $\Lambda$-CDM cosmology with $H_0=67.9\,{\rm km}\,{\rm s}^{-1}{\rm Mpc}^{-1}$ and $\Omega_{\rm m}=0.3065$~\cite{Planck:2015fie} is assumed to compute the redshift, as well as the prior over the luminosity distance. 
In the npE sector, we consider both $(\zeta_1,\zeta_2)$ and $(\zeta_b,\varphi)$ when parametrizing the model and, in each case, we choose a uniform prior over the two parameters within the unit circle. 

In order to estimate the individual-event posteriors, we perform nested sampling using \texttt{Bilby} with the \texttt{dynesty} sampler. 
Each parameter estimation run uses 1000 live points and stops at \texttt{dlogz=0.1}. The MCMC evolution in each nested sampling step is done with the \texttt{Bilby}-implemented \texttt{acceptance-walk} method, with evolution length controlled by \texttt{naccept=60} when the GR base waveform is \texttt{IMRPhenomPv2} or \texttt{naccept=100} when the GR base waveform is \texttt{IMRPhenomXPHM}.
We have checked that our individual inference runs are robust to these sampler choices. 
We first apply the nested sampling to the npE analysis assuming the $(\zeta_1,\zeta_2)$ parametrization. Then, we reweight the sample by $1/\sqrt{\zeta_1^2+\zeta_2^2}$ to estimate the alternative npE posterior assuming the $(\zeta_b,\varphi)$ parametrization.  

For the hierarchical inference, we reuse the posterior sample from each individual-event npE analysis assuming the $(\zeta_1,\zeta_2)$ parametrization.
Specifically, we use each posterior sample to fit a Gaussian kernel density estimation (KDE), where we adopt the \texttt{scipy}~\cite{2020SciPy-NMeth} implementation of the KDE and set the bandwidth following Scott's rule~\cite{scott2015multivariate}.
Because the prior in the npE sector is flat and the likelihood is invariant under parameter transformation, we have
\begin{align}
    &\mathcal{L}_{\rm npE}^{(i)}(s^{(i)}|\zeta_b^{(i)},\varphi^{(i)}) \notag\\
    &\;= \mathcal{L}_{\rm npE}^{(i)} \big(s^{(i)} \big| \zeta_1^{(i)}=\zeta_b^{(i)}\cos\varphi^{(i)},\zeta_2^{(i)}=\zeta_b^{(i)}\sin\varphi^{(i)} \big) \notag\\
    &\;\propto p_{\rm npE} \big(\zeta_1^{(i)}=\zeta_b^{(i)}\cos\varphi^{(i)},\zeta_2^{(i)}=\zeta_b^{(i)}\sin\varphi^{(i)} \big| s^{(i)} \big). \label{eqn:likelihood_kde}
\end{align}
We use the above relation to approximate the individual-event npE likelihood function with the posterior KDE and further simplify the hierarchical likelihood following Eq.~\eqref{eqn:likelihood_hierarchical_gaussian}, where $K^{(i)}$, $w_j^{(i)}$, $\vec{\mu}_j^{(i)}$, and $C_j^{(i)}$ take the corresponding values of the KDE.
We choose a prior uniform over $\mu\in[-1,1]$, $\sigma\in[0,1]$, and $\bar{\varphi}\in[0,\pi]$, and use \texttt{Bilby} again with the same settings for individual-event analysis to sample over these three parameters of the hierarchical model. 

One potential flaw in Eq.~\eqref{eqn:likelihood_kde} is that the posterior KDE may fail to represent the likelihood due to the limited range of our prior. As seen in Fig.~1 of the main text, a few individual-event npE posteriors are sharply cut at the unit circle, which is a clear artifact of the npE prior and the likelihood support is expected to continuously extend beyond that point. 
However, if the most of the weight of the individual-event posteriors is around the GR point, the combination through hierarchical inference should be able to suppress such an artifact near the unit circle, and the existence of only a few problematic individual-event posteriors should not significantly impact the final result. 
On the other hand, if there are too many prior-affected events, the hierarchical inference should also return a population distribution that tends to rail against the individual-event prior boundary at the unit circle.
In Fig.~2 of the main text, however, we have shown that the reconstructed $\zeta$ population dies off way before reaching the unit circle boundary, proving that the prior impact is low. Therefore, we conclude that Eq.~\eqref{eqn:likelihood_kde} remains an appropriate approximation for our purposes in this paper.

Another concern is that the na\"ive Gaussian KDE can extend outside the individual-event npE prior range, and, hence, affect the numerical result. To address this problem, we have run a diagnostic analysis where both the Gaussian KDE and the population model are truncated at $\vec{\zeta}=1$. We have checked that the posteriors of our hierarchical inference does not change under the truncation, implying that the na\"ive Gaussian KDE applied here is numerically robust.

\subsection{Posteriors from individual events}
\begin{figure*}
    \centering
    \includegraphics[width=0.98\linewidth]{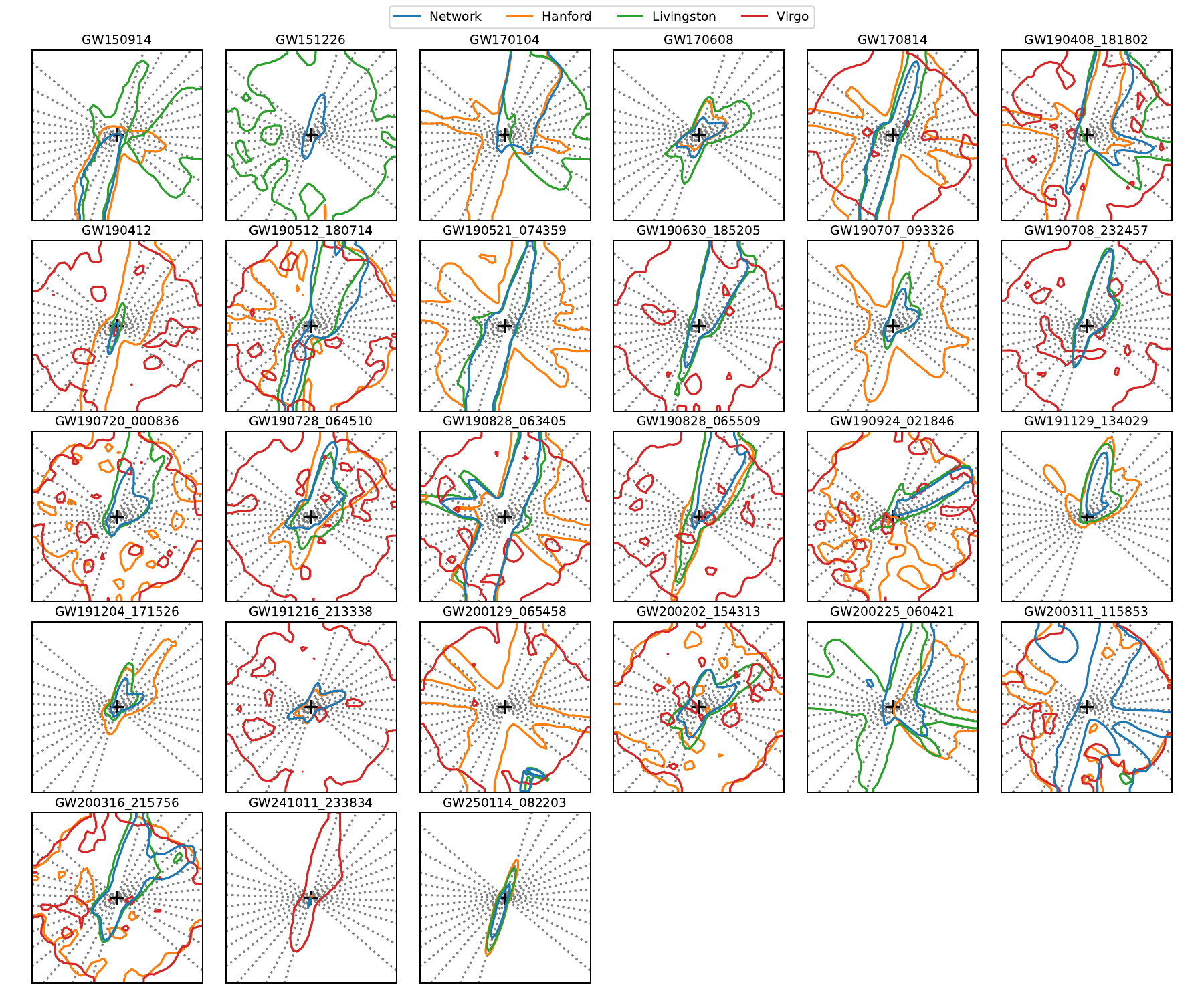}
    \caption{Posterior 90\% credible contours from individual-event npE analysis assuming a uniform prior over $\zeta_1$ and $\zeta_2$ within the unit circle. Each panel presents the analysis result of an event in Table~\ref{tab:events}, as suggested by the panel title, and apart from the posterior generated using data of the entire network (blue), there are also posteriors generated using data of individual detectors including the Hanford detector (orange), the Livingston detector (green), and the Virgo detector (red). 
    See Table~\ref{tab:events} for the list of detectors used in each event.
    The ticks and labels of the panel axes have been omitted for simplicity. For all panels, the $x$-axis represents $\zeta_1\in[-1,1]$ and the $y$-axis represents $\zeta_2\in[-1,1]$. Each panel is also gridded by gray dotted lines, which are the same PN lines shown in Fig.~2 in the main text. The black ``$+$'' at the center of each panel marks GR.}
    \label{fig:individual_event_posteriors}
\end{figure*}

Figure~\ref{fig:individual_event_posteriors} shows the posteriors from the individual-event npE analyses, assuming uniform priors over $\zeta_1$ and $\zeta_2$. 
For each event, we present the 90\% credible contours of the marginalized posteriors in the $\vec{\zeta}$ plane, generated with the analysis results using data of the entire network (black) and individual detectors, including the Hanford detector (blue), the Livingston detector (orange), and the Virgo detector (green), respectively.

Let us now check whether the npE analysis of each event has been affected by any noise artifacts. This can be done by comparing the $\vec{\zeta}$ posteriors between different detectors. Because noise artifacts are detector-specific and should not be shared across the network, any inconsistency between these $\vec{\zeta}$ posteriors raises a red flag. 
In our case, we look for events where the 90\% credible contours obtained with different detector networks are incompatible with each other, and we identify three events that have been likely affected by noise artifacts: GW151226 (between the network posterior and the Hanford posterior), GW200129\_065458 (at least between the Hanford posterior and the Livingston posterior), and GW200311\_115853 (at least between the Hanford posterior and the Livingston posterior). 
Among the three events, GW200129\_065458 is already known to have a glitch-removal artifact in the LVK open data that significantly impacted the inference of spin precession assuming GR~\cite{Payne:2022spz}.
Therefore, we consider the above diagnosis robust and exclude the three events listed from the results presented in the main text.

For the rest of the events, the network posteriors are mostly consistent with GR by enclosing $\vec{\zeta}=0$ inside the black 90\% credible contours. Exceptions are GW191129\_134029 and GW190924\_021846, but their network 90\% credible contours are not far from the GR point. There are also a few critical events such as GW150914, for which GR appears to be sitting on the 90\% posterior contour. 
However, these cases should be seen as affected by instrument noise that happened to behave like non-GR deviations. 
This argument has been further strengthened by the hierarchical npE test result, from which we have presented a $Q_{\rm GR}=34\%<90\%$ in the main text.
Here, the consistency with GR at the population level can be interpreted as a result from averaging out the random noise effects of individual events.

\subsection{Posterior from hierarchical inference}
\begin{figure}[htbp]
    \centering
    \includegraphics[width=0.98\linewidth]{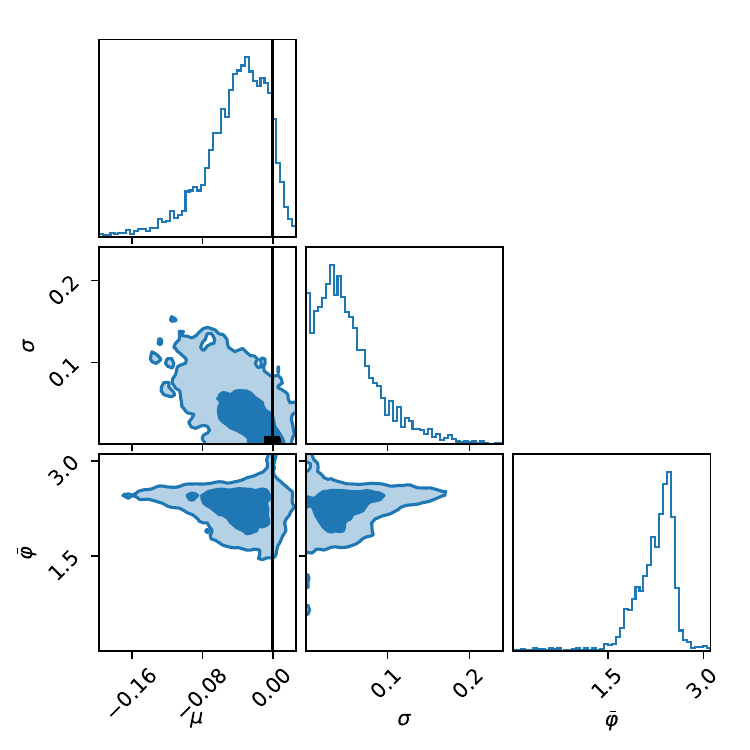}
    \caption{
    Hierarchical posterior of the population hyperparameters. The blue contours enclose the 50\% and 90\% credible regions, respectively. The black dot and lines mark GR ($\mu=0=\sigma$).
    }
    \label{fig:hierarchical_posterior}
\end{figure}

Figure~\ref{fig:hierarchical_posterior} shows the posterior of the population hyperparameters $p_{\rm h}(\mu,\sigma,\bar{\varphi}|\{s\})$ from hierarchical inference. 
Observe that GR ($\mu=0=\sigma$) is well enclosed by the 90\% credible contour in the $\mu$--$\sigma$ plane.
The credible level at which GR is critically enclosed, namely the GR quantile, is accessible through Eq.~\eqref{eqn:gr_quantile} and our estimate yields $Q_{\rm GR}=0.34$ as reported in the main text. 
Furthermore, the hyperparameters posterior leads to the reconstructed population of $\vec{\zeta}$ through Eq.~\eqref{eqn:pop_recon}, which we present in Fig.~2 in the main text.

\end{document}